\def\PRL{{Phys. Rev. Lett.} }
\def\PRA{{Phys. Rev.} A }

\documentclass{ws-ijqi}

\begin{document}

%%%%%%%%%%%%%%%%%%%%% Publisher's Area please ignore %%%%%%%%%%%%%%
\catchline{}{}{}{}{}
%%%%%%%%%%%%%%%%%%%%%%%%%%%%%%%%%%%%%%%%%%%%%%%%%%%%%%%%%%%%%%%%%%%

\title{Recent experiments performed at
"Carlo Novero" lab at INRIM on Quantum Information and Foundations
of Quantum Mechanics.  }

\author{ G. BRIDA, N. ANTONIETTI,
M. GRAMEGNA, L. KRIVITSKY, F. PIACENTINI, M.L. RASTELLO, I. RUO
BERCHERA, P. TRAINA, M. GENOVESE \footnote{genovese@inrim.it}.}

\address{INRIM; strada delle Cacce 91, 10135 Torino, Italy}

\author{E. PREDAZZI}

\address{Dip. Fisica Teorica, Univ. Torino, via P. Giuria 1, 10125 Torino, Italy }

\maketitle

\begin{history}
\received{30-05-2006}
%\accepted{Day Month Year}
%\comby{(xxxxxxxxxx)}
\end{history}

\begin{abstract}
In this paper we present some recent work performed at "Carlo
Novero" lab on Quantum Information and Foundations of Quantum
Mechanics.
\end{abstract}


\section{Introduction}    %) A SECTION HEADING

"Carlo Novero" laboratory (named after our bewailed colleague and
friend who founded this activity in our institute) is a facility
at the Italian National Institute of Metrological Research (INRIM)
devoted to the experimental study of the foundations of quantum
mechanics and quantum information by using quantum optical states.

In particular most of the activity was addressed to produce and
use Parametric Down Conversion (PDC) biphoton pairs \cite{nosrev}.

A first application of these states concerned quantum metrology,
i.e. the calibration of single photon detectors \cite{det}, but
the same states also find wide application to the related fields
of Quantum Information and Foundations of Quantum Mechanics: in
this paper we present our most recent and interesting studies on
these two subjects.

\section{Two type I  crystals source and application to tests of realistic theories}

In our laboratory we used various different sources of
polarization entangled photons,  some based on a type II PDC
crystal other on the superposition of the emission of two type I
PDC crystals.

A large part of our experimental work concerning tests of local
realism was realized with the last one.

\begin{figure}[h]
\includegraphics[width=8.5truecm,scale=1.2]{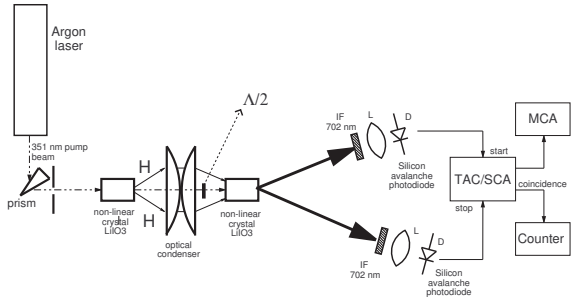}\caption{Sketch of the bright source of polarization
entangled photons realized by superimposing
 two type I PDC emissions. The detection apparatus is shown as well.}%
\label{protocol0}%
\end{figure}

This source (Fig. \ref{protocol0}) was built by superimposing the
emission of two separated type I crystals \cite{na} ($LiIO_3$, 1.5
cm length), whose optical axes were at $90^o$, by using an optical
condenser. The pump polarization was rotated of $90^o$ between
them. This setup realized a very bright source \cite{nosrev} of
polarization entangled states (10 kHz for the coincidence rate at
200 mW pump power) and a very good superposition was possible (in
principle much better than what realizable with two adjacent thin
crystals \cite{kw}), therefore representing an interesting
resource for quantum information protocols.

A first application was to test Bell inequalities \cite{pr} with
non-maximally entangled states \cite{nosww}, a step toward a
solution of detection loophole, since the quantum efficiency limit
for a loophole free experiment is lowered for non maximally
entangled states \cite{eb}. A clear violation of Clauser-Horne
inequality (less than zero for local realistic theories) was
observed, $513 \pm 25$.

\begin{figure}[h]
\includegraphics[height=7cm]{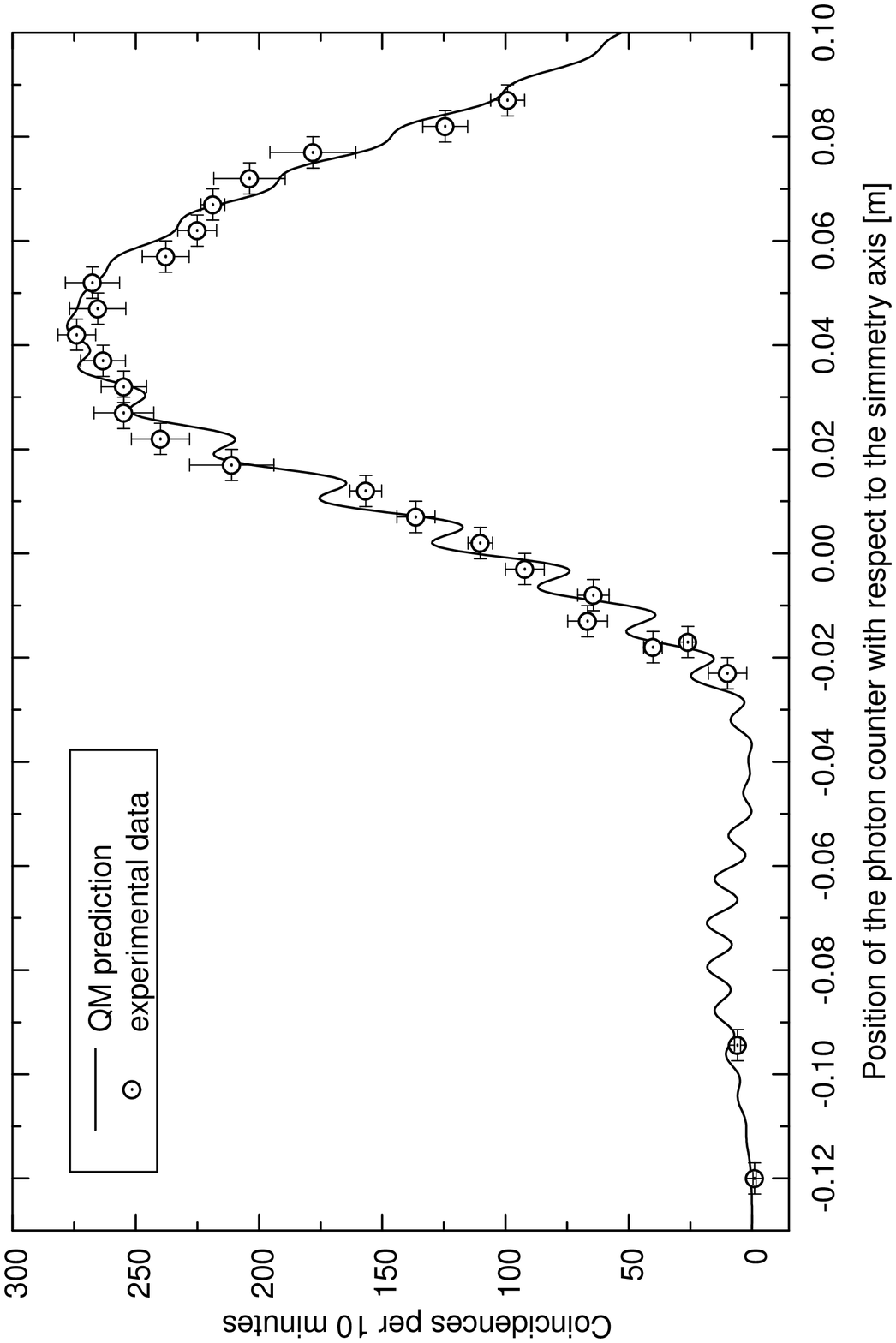}\caption{ Coincidences data compared with quantum
mechanics predictions (solid curve). On the x-axis we report the
position of the first detector with respect to the median symmetry
axis of the double slit.
 The second detector is positioned at -0.01 m (out of scale). The leftmost region of the data is inaccessible since the
 two detectors overlap, while on the right, a rather flat behavior for coincidences is
 predicted. A clear coincidence signal in the forbidden region for
 dBB  is observed (negative part of x axis). }
\label{protocol1}%
\end{figure}

\begin{figure}[h]
\includegraphics[angle=-90,width=7cm]{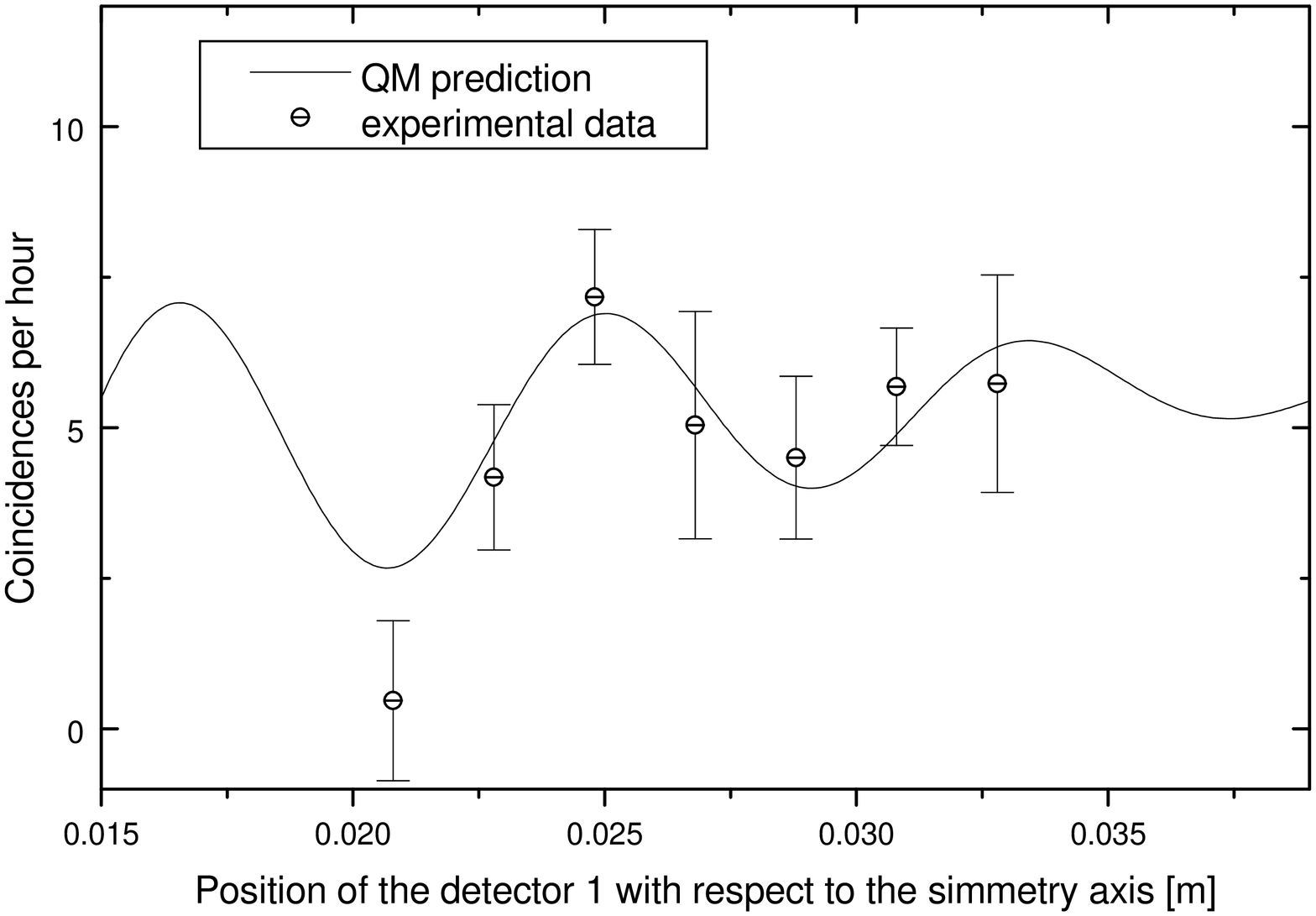}
\caption{Coincidences data (with a 2 mm iris) compared with quantum mechanics
predictions (solid curve) for the 4th order interference.
 On the x-axis we report the position of the first detector with respect to the median symmetry axis of the double slit.
 The second detector is positioned at -0.01 m (out of scale). The leftmost region of the data is inaccessible since
 the two detectors overlap, while on the right, a rather flat behavior for coincidences is
predicted.}%
\label{int}%
\end{figure}

\begin{figure}[h]
\includegraphics[width=8truecm]{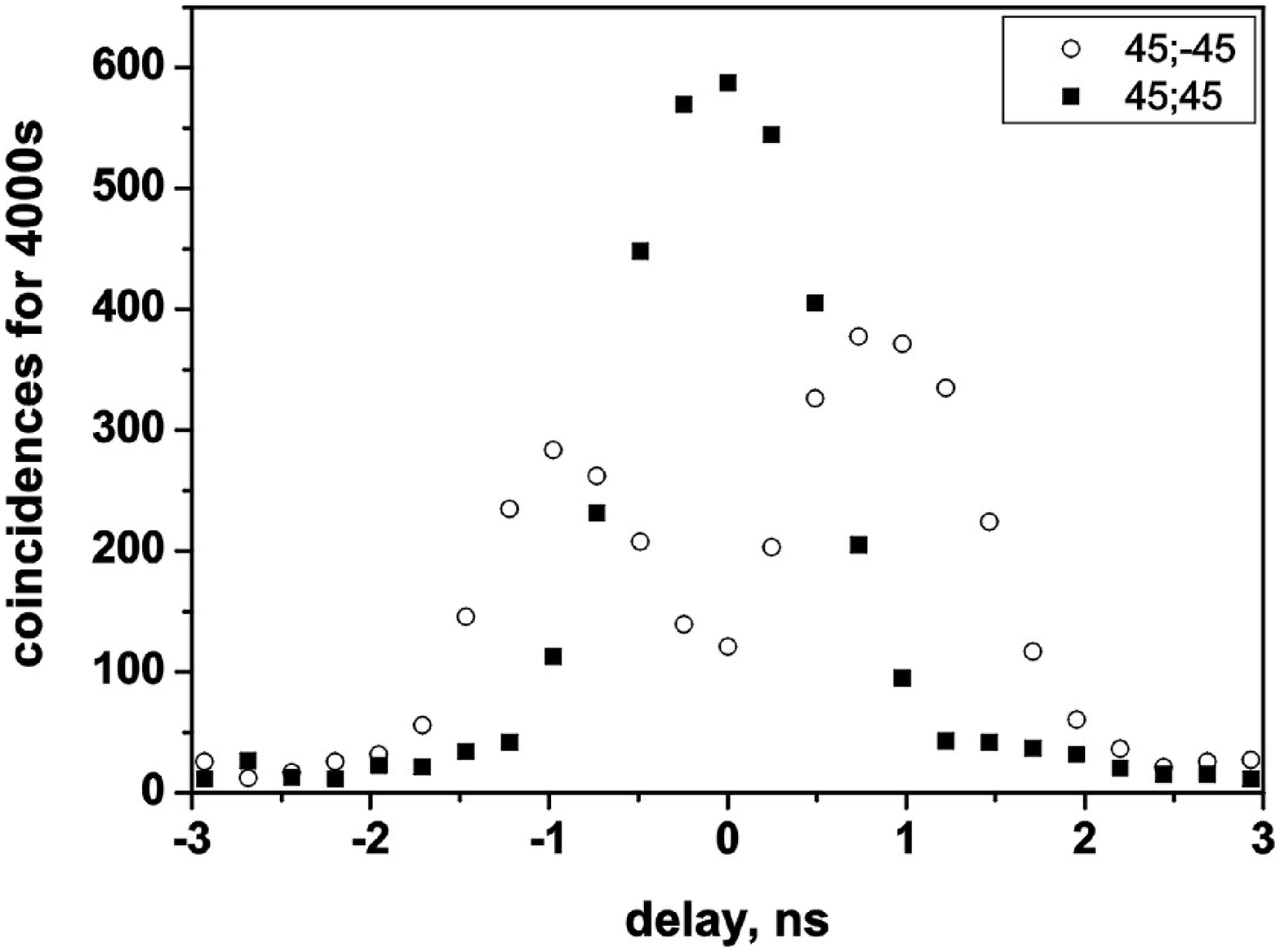}\caption{Coincidences
observed when the polarizers are parallel or orthogonal in $45^o$
basis. The configuration is the double fiber cross with a Faraday
mirror reflecting back the biphotons and a 250 m fiber.
}%
\label{protocol2}%
\end{figure}

Furthermore, this experiment allowed a clear negative test of
stochastic electrodynamics \cite{sed}, a theory built for
reproducing quantum electrodynamics results in a classical field
theory framework where a zero-point field is introduced. In its
subpart concerning the quantum properties of radiation, named
stochastic optics, it was forecasted that Bell inequalities should
not be violated under a certain level of detection rate
\cite{cas}. Indeed a clear violation of Clauser-Horne inequality
was observed in our experiment even being well (many order of
magnitude) under this threshold.

Incidentally, as a further test of stochastic optics we also
searched for a spontaneous up conversion (SPUC) emission predicted
in this theory \cite{puc}. More in details we pumped with both a
diode laser at 789 nm (50 mW power) and a Neodimium-Yag laser beam
(1064 nm, 0.51 W power) a 1.5 cm Lithium Iodate crystal in the
configuration were a stimulated emission was emitted when a UV
pump (351 nm Argon laser beam) was present. In the same
configuration SPUC was expected when the UV beam was turned off.
We did not observe any emission by monitoring the emission after
the crystal with a  ccd camera \cite{nosW} (i.e. SPUC signal was
at least 160 times smaller than the PDC one). Again, no emission
was observed (by scanning substantially all the possible angles
for the  emission) when the same experiment was reproposed by
using a 5 mm BBO crystal pumped by a 789 nm wave length, 90 mW
power, diode laser beam.

Thus, altogether all these negative results clearly falsify this
theory \footnote{Incidentally, also experiments at single photon
level where zero-point field does not look to play a relevant role
\cite{wpd} appear  not to be describable in such a theory.}.

Finally, we would like to mention that, by substituting the second
crystal with a double slit, it was possible to realize the
experiment proposed by ref. \cite{parta} for testing standard
quantum mechanics (SQM) against de Broglie-Bohm theory. In extreme
synthesis, ref. \cite{parta} proposed that when two identical
bosonic particles cross each a slit of a double slit at the same
time they never cross the symmetry axis of the slits at variance
with SQM predictions. Opposite to this prediction we clearly
observed  coincidences of identical photons (702 nm PDC conjugated
photons) in the same semiplane \cite{dBB} after crossing each a 10
micrometers slit (being the two slits separated of 100
micrometers), see Fig. \ref{protocol1}. As a further result
obtained with this set-up in Fig. \ref{int}  a detail of the
coincidence curve is shown: 4th order interference is clearly
observed (rejecting the absence of interference at 95\% confidence
level both for a $\chi^2$ and a run test). On the other hand, 2nd
order interference is not observed since in this case the
distinguishability of the path of the two photon  is kept.

\section{Type II PDC sources and applications}

Various other experiments were realized with type II PDC sources.

Here we would like only to mention the two most recent of them.

In the first one we generated collinear degenerate biphotons
\cite{fib} that travelled through a dispersive medium (a fiber)
before being split by a beam splitter and detected (see Fig.
\ref{set}). The temporal growing of the wave packet into the fiber
was demonstrated to allow on the one hand to measure interference
effects \cite{mas} otherwise under temporal resolution of detector
apparatuses and on the other hand to restore indistinguishability
(by erasing longitudinal walk-off) and therefore entanglement for
pairs in the center of the coincidence peak (allowing the
observation of a violation of Bell inequalities). The effect was
studied for various configurations: 250 m and 1 km fiber and two
passes through a 250 m fiber with a Faraday mirror reflecting back
the light (since the Faraday mirror acts as a time reverse operation
\cite{far} this configuration allowed for an erasure of fiber
polarization effects allowing a very high stability of the setup).
The Full Width at Half Maximum of the coincidence peak grew from 0.8
ns in absence of the fiber up to 4.5 ns with the 1 km one. The high
visibility in the center of the peak (see Fig. \ref{protocol2})
between the configuration with polarizers orthogonal and parallel in
$45^o$ basis (where the pump beam is vertical) certifies the
restoration of the entanglement. Indeed, for example, with the 1 km
fiber a clear violation of the Bell inequality \begin{equation}
R={|N(\pi / 8) - N(3 \pi /8)| \over N(\infty,\infty) } \leq 0.25
\label{Rsym} \end{equation} ($N(\theta_1,\theta_2)$ being the number
of coincidences for polarizers settings $\theta_1,\theta_2$ and
$\infty$ denoting no selection) was observed when selecting the
central 0.43 ns, $R= 0.322 \pm 0.061$.

This setup also allowed the study of transmission in fiber of
polarization entangled states, permitting researches on
decoherence effects. The rapid variation of the polarization
effects suggested a scheme for realizing a controlled decoherence
on a quantum channel based on plunging the fiber into an
ultrasound bath. This setup is now under realization and will find
applications ranging from characterization of the channel as a
Completely Positive map to studies on decoherence in realistic
Quantum Key Distribution protocols.

\begin{figure}[h]
\includegraphics[width=8truecm]{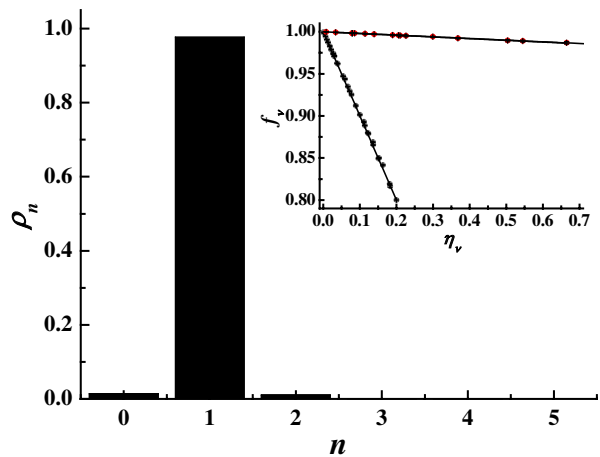}\caption{Reconstruction of the  photon distribution
for the heralded single-photon state produced in type II PDC.
Inset: The steepest curve corresponds to experimental frequencies
$f_{\nu}$ of no-click events as a function of the quantum
efficiency $\eta_{\nu}$ for a PDC heralded photon state compared
with the theoretical curve $p_\nu = 1-\eta_\nu$. The small vacuum
and two photon components are in agreement with what estimated.
For the sake of completeness the curve for data on a weak coherent
state is shown as well (highest one).
}%
\label{distribution}%
\end{figure}

The second experiment was addressed to reconstruct the photon
statistics by using on/off detectors only \cite{mat}. Since the
knowledge of the statistics of quantum optical states is a
prerequisite for various experiments ranging from quantum optics
to quantum information, and no available detector can well
determine the number of incident photons, this research has large
relevance for a wide spread application. In little more detail, we
reconstructed the photon statistics of various optical states by a
maximum likelihood algorithm applied to data obtained by varying
the optical transmittivity through the introduction of calibrated
neutral filters, both for mono-partite \cite{mono} and bi-partite
\cite{bi} cases, showing the large potentialities of the method.
In particular, for example, very interesting results were obtained
with single heralded photons produced by type II PDC (see Fig.
\ref{distribution}).

\section{Acknowledgements}
This work has been supported by MIUR (FIRB RBAU01L5AZ-002 and
RBAU014CLC-002, PRIN 2005023443-002 and 2005024254-002), by
Regione Piemonte (E14), and by "San Paolo foundation".

We would like to thank Maria Chekhova, Matteo Paris, Andrea Rossi,
Maria Bondani, Guido Zambra, Alessandra Andreoni and Partha Ghose
for the fruitful collaboration and pleasure of working together
during the realization of some of the experiments described in
this review.

\end{document}